\documentclass[preprint]{icarus}
\usepackage{graphicx}
\begin{document}

\begin{center}

Title of the paper: 

Photometry of comet 9P/Tempel 1 during the 2004/2005 approach and the Deep
Impact module impact

\bigbreak\bigbreak

Authors: 

G.A. Milani$^{1,2}$,
Gy. M. Szab\'o$^{1,3,4,+}$,
G. Sostero$^5$,
R. Trabatti$^6$
R. Ligustri$^7$,
M. Nicolini$^8$,
M. Facchini$^8$,
D. Tirelli$^9$,
D. Carosati$^10$,
C. Vinante$^{1,11}$,
D. Higgins$^{12}$

\bigbreak\bigbreak


$^{1}${CARA, Unione Astrofili Italiani, vicolo Osservatorio 2, 35122 Padova, Italy}

$^{2}${University of Padova, Dept. of Biomedical Exp. Science, viale G. Colombo 3, 35121 
Padova, Italy}

$^{3}${University of Szeged, Dept. of Experimental Physics, D\'om t\'er 9, H-6720
Szeged, Hungary}

$^{4}${Magyary Zolt\'an Postdoctoral Research Fellow}

$^{5}${Remanzacco Observatory, via S. Stefano, 33047 Remanzacco, Italy}

$^{6}${Stazione Astronomica ``Descartes'', via Lambrinia 4,  27013 Chignolo P\'o , Italy}

$^{7}${Talmassons Observatory, via Cadorna, 33030 Talmassons, Italy}

$^{8}${``G. Montanari'' Observatory, via Concordia 200, 41032 Cavezzo, Italy}

$^{9}${Tirelli's Observatory, 36040 Sossano, Italy}

$^{10}${Armenzano Astronomical Observatory, 06081 Assisi, Italy}

$^{11}${Associazione Astronomica Euganea, via N. Tommaseo 70, 35131 Padova}

$^{12}${Hunters Hill Observatory, 7 Mawalan Street, Canberra, Australia}

$^+$e-mail: szgy@titan.physx.u-szeged.hu

\end{center}

25 pages (including cover pages), 9 figures, 1 table

\newpage

\begin{center}

Proposed runnung head:

Photometry of Tempel 1 during the 2004/2005 approach

\bigbreak\bigbreak\bigbreak

Name and address to which editorial correspondence and proofs should be directed:

\medbreak

Gyula M. Szab\'o

University of Szeged

Department of Experimental Physics and Astronomical Observatory

D\'om t\'er 9.

H-6720 Szeged, Hungary

\medbreak

e-mail: szgy@titan.physx.u-szeged.hu

\end{center}

\newpage

\begin{center}

Abstract

\bigbreak\bigbreak

The results of the 9P/Tempel 1 CARA (Cometary Archive for Amateur Astronomers) observing campaign is presented. The main goal was to perform an extended 
survey of the comet as a support to the Deep Impact (DI) Mission.
CCD R, I and narrowband aperture photometries were used to monitor the $Af\rho$ quantity.
The observed behaviour showed a peak of 310 cm 83 days before perihelion, but we argue that it can be distorted by the phase effect, too.
The phase effect is roughly estimated around 0.0275 mag/degree, but we had no chance for direct determination because of the
very similar geometry of the observed apparitions. The log-slope of $Af\rho$ was around $-0.5$ between about 
180--100 days before the impact but
evolved near the steady-state like $0$ value by the impact time. The DI module impact caused an about 
60\%{} increase in the value of $Af\rho$ and a cloud feature in the coma profile which was observed just after the event. 
The expansion of the ejecta cloud was consistent with a fountain model with initial projected velocity of 0.2 km/s and 
$\beta$=0.73. Referring to a 25~000 km radius area centered on the nucleus, the total cross section of the ejected dust 
was 8.2/$A$ km$^2$ 0.06 days after the impact, and 1.2/$A$ km$^2$ 1.93 days after the impact ($A$ is the dust albedo). 
5 days after the event no signs of the impact were detected nor deviations from the expected activity referring both to the 
average pre-impact behaviour and to the previous apparitions ones.

\end{center}


\section{Introduction}

Comet 9P/Tempel 1 is a well known periodic comet of the Jupiter family ($5.4$ years period, q = $1.4$ A.U.) that librates around a mean motion 
resonance (Fernandez, 2005) and  undergoes close encounters with our planet. 
It  was selected as the Deep Impact mission (DI hereafter) target. After a successful launch (early 2005) the spacecraft had a close 
encounter with 9P/Tempel 1 on July 4th. The main module performed a flyby with a $400$ km minimum distance from the nucleus, while the 
impactor module hit the surface, excavated a crater and produced an expanding ejecta cloud (A'Hearn et al, 2005).

Previous observations collected by DI showed that this comet displayed short term events (jets in the coma, outbursts). 
Extended monitoring of this comet was therefore highly desired in order to discern the usual variations from the effects 
of the impact and the variations related to the normal activity of the comet. In order to ensure a continuous monitoring, a 
call for observations  was made by DI to amateur astronomers, and a Small Telescope Scientific Program (McLaughlin et al. 2004) 
was developed to collect images and data.

The CARA project (Cometary Archives for Amateur Astronomers, http://cara.uai.it/) 
was established in  2003 to provide continuous photometric monitoring of specific comets, and in 2004 we joined this
call for observations. The observers are skilled amateur astronomers with moderately small telescopes (typically with aperture between $20$ and $60$ cm) 
who already possesses the necessary knowledge in image reduction and data analysis. By the end of  CARA collected more than 
$3000$ data points for almost $30$ comets. Our observations cover 10 months , and thanks to the contribution of amateur
astronomers, the network was able to collect data on almost every clear night.
A basic description of this network, the data processing and the observational circumstances is introduced in Section 2.
The results are summarized in Section 3 while we conclude in Section 4.

\section{Observations}

{\bf Fig. 1 comes here}

The aim of our campaign was to obtain photometric data to derive the $Af\rho$ quantity (A'Hearn 1984). 
This quantity measures the dust continuum and allows a comparison of data obtained at different sites, epochs, 
geometrical circumstances, and/or with different telescopes and photometric apertures. It is the product of Bond-albedo $A$ 
(Bond, 1861, Bell, 1917), the $f$ filling factor of the grains within the aperture and $\rho$ as the radius of the field of view at the comet.
\begin{equation}
Af = \bigg({2DR \over \rho} \bigg)^2 \cdot {F_{com}\over F_{sol}}.
\end{equation}
Where 
the Earth-comet distance $D$ and $\rho$ is in cm, the Sun-Earth distance $R$ is in AU, $F_{sol}$ is the flux of the Sun at 
$1$ AU referring to the photometric band used, $F_{com}$ is the observed flux from the comet.

If one assumes an isotropic dust ejection from a point source, and if dust moves with constant velocity $v_e$, 
and effects of solar radiation pressure are neglected, the expected column density of dust is  $\pi Q /(4v \rho)$. Here $Q=dN/dt$ 
is the dust production rate. In this case the coma is expected to have a $1/\rho$ photometric profile, that is 
why its $Af\rho$ is independent of $\rho$. 
The $\Sigma$ total cross section of dust is 
\begin{equation}\Sigma = {\pi \over 4A}\ Af\rho \cdot \rho .\end{equation}

$Af\rho$ is approximately independent of $\rho$ for the majority of the active comets and for a wide range of apertures 
(e.g. between $5~000$--$50~000$ km radius). 
The average profile of the coma brightness is however better described with a power-law as 
$\rho^\gamma$, where gamma is the slope calculated by the $\log rho$ dependence of 
$\log Af\rho$. Denoting the various coefficients simply with $K$ and $K'$, 
$\log Af\rho = K + \log \left( {1\over\rho} \int \rho\rho^\gamma {d}\rho \right) = K' + (\gamma+1)\log\rho,$
consequently 
\begin{equation}
{d \log Af\rho \over  d \log \rho} = \gamma+1.
\end{equation}
$Af\rho$ describes the isotropic structures, the non-isotropic features in the coma can be better investigated with image processing. 
Even if $Af\rho$ is almost independent of $\rho$, one cannot immediately conclude 
that the comet is ejecting dust with constant speed or with perfect isotropy. Farther from the nucleus as the steady state is no 
longer valid, $Af\rho$ begins decreasing.

For the photometry of dust CARA suggests the use of CCD and R, I (Bessel, 1990) or a narrowband filters (e.g. $647$ nm / $10$ nm, 
Rc hereafter). In Fig. 1 the filter transmission curves are compared with an indicative cometary spectrum (adapted from Lamy 1986). 
The Rc filter is used as an affordable cheap alternative to the standard HB (Farnham et al., 2000) and IHW 
(Edberg, 1983) dust continuum sets. The bandpass of the Edmund $647/10$ nm filter used at different sites was 
verified with an Aminco DW 2000  spectrophotometer at the Biomedical Experimental Sciences Department of the Padova University.
The characteristics are very close to what the manufacturer provided, with no appreciable blue or infrared-leak. 
The Rc filter had been calibrated with the S band of the Vilnius System (Kakaras, 1968), a procedure that 
was successfully applied for the photometry of the tail of C/1996 B2 Hyakutake (Fulle et al., 1998).
  
As a suitable S/N has to be achieved, in fainter comets the 
narrow-band filters would lead to very long exposures. In this case the usability 
of the Rc filter is limited and R and I filters are preferred. Unfortunately R and I
filters involve some molecular emissions bands, mainly NH$_2$ and C$_2$ (R) and CN (I).
That is why the $Af\rho$ can be overestimated in some comets, but
in the case of dusty objects the magnitude of this excess is about $10$\% or less
(Stern et al., 1999). The other advantage is that the effective wavelength of R and Rc filters are nearly coincident, 
allowing an easy comparison of the data as well as a check on the possible emission line contamination in the $R$ band.
For Tempel 1 the R and Rc data are in good agreement which indicates fairly negligible gas contamination.

Solar constants used for $Af\rho$ calculations are:
$V_\sun$  = $-$26\fm71	(Caldwell et al., 1993),
$(B-V)_\sun$ = 0\fm66 	(Boyle et al., 1998),
$(V-R)_\sun$ = 0\fm36,
$(V-I)_\sun$ = 0\fm69	(Hardorp et al., 1978) and
$(V-S)_\sun$= 0\fm59    (Straizys \&{} Valiuaga, 1994)

In order to examine the structure in more details and to better the mass-loss
estimates a Surface Brightness Profile Analysis has been done. We determined
the slopes of the $\log Af\rho$--$\log \rho$ profiles from each single observations. The linear
regression was fitted between $\log \rho=3.8$ and $4.8$, where all the profiles were
linear. Finally we checked the linearity of the profiles and the quality of the fits 
for each observations individually. 

\subsection{The source of data - CARA observing campaign}

{\bf Table 1. comes here}

{\bf Fig. 2 comes here}

$26$ observers were participating the $9$P/Tempel 1 observing campaign, 
providing a set of about $800$ aperture photometry measurements. $473$ and $225$
raw data points were collected pre-impact and post-impact, respectively, and $86$
points fall $\pm3$ days within the time of impact.
In every observing run, each observer collected a sequence (usually dozens) of images of the comet, 
together with $10$--$20$ dark frames and flat fields 
(mainly sky flats in twilight or sometimes dome flats). A master dark-frame and flat-field were obtained from 
the average of the previous sets. Each comet image was individually pre-processed with the master-dark 
and the master-flat frames. We also checked that no random background variation (or noise) occurred in the 
image sequences that would question the reproducibility of the measurements.
The pre-processed images were then aligned on the comet and averaged.  
Images with bright stars superimposed to the coma were not considered for the analysis.

The participants, observing sites, telescopes and filters are summarized in Table 1.
The telescope sizes of our network range from small $12$ cm  telescopes up to
$60$ cm reflectors, the image scale is typically around $1$--$2$ arcsec/pixel.

$Af\rho$ was measured with aperture photometry, centering the
apertures at the photometric center of the comet. In order to describe the coma profile,
at least three measuring windows are required. According to the primary 
recommendation of CARA guidelines they should be close to $50~000$, $25~000$ 
and $12~500$ km radius. However, in some cases (e.g. a comet at small geocentric distance 
or with a coma not extending too far from the nucleus) smaller apertures 
(sub-multiples of the standard  windows) are suggested, too. 
In order to avoid under-sampling in measurements the lower limit for the smaller 
adopted aperture size is three times the seeing value. 
A sample pair of images is shown in Fig. 2.

Comparison stars were selected in the field or, if this was not possible, at similar air-mass within 3 degrees from the comet. 
The use of Landolt photometric stars was rarely considered as they seldom occurred close to the target, and the most sites
had rarely photometric sky for reliable transformations. This is why other reference stars were selected.
As a reasonable compromise the Hipparcus and Tycho Main catalogues were used as they offer 
an average accuracy close to 0.01 magnitudes in the Johnson System (Perryman, 1997). The catalogues also provide 
information about the variability of a large number of stars. In few cases Tycho stars with worse accuracy had to be used (within $0.05$ mag).
Comparison stars had to have a colour index around the solar value ($0.4< B-V <0.8$).
Using solar like stars the colour term has a negligible effect in the transformations (within $2$--$3$\%{}), and under
non-photometric conditions it may be dispensed.

The measurement of the surface brightness profile consisted of determining $d \log Af\rho /d \log\rho$ (Eq. 3). Between $7000$
and $35~000$ km radius, the $\log(Af\rho)=M\cdot\log(\rho)+N$ line was fitted to the raw observations via $M$ and $N$. 
All points had the same weight. The resulting $A$ was accepted as the logarithmic profile parameter, while the width of
confidence interval was accepted as the uncertainty. In this step we considered only the images with R or I filters.
The logarithmic profile parameter does not depend on the exact value of $Af\rho$, so it is not affected by
the photometric calibration. The error sources are the background subtraction and the faint stars in the coma,
therefore we double-checked the used images were of acceptable quality.
 
\subsection{Data analysis}

In order to ensure a high consistency of data, to avoid systematics or other discrepancies between the observers, 
a specific software ({\sc Winafrho, Xafrho} respectively for Windows and Unix based computers) had been developed 
and supplied for free to the observers. Aperture photometry was performed by means of square apertures. The code then re-estimated
the measures referring to a circular window taking into account the $1/\rho$ gradient for a theoretic steady state coma. 
The systematic differences introduced by square 
windows compared to circular ones are estimated to be less than 5\%{}
even in the case of very asymmetric comets like 19P/Borrelly (Milani, 2005).
Sky background is computed from a manually selected image area that is close to the comet but is far enough
from the coma, and contains no stars. The sky value is set by the software as the lowest value of the median of 
three selected image columns.
This procedure allowed to exclude the contamination from faint stars, cool and hot pixels and/or cosmic rays.

The calculation of $Af\rho$ includes the geometrical circumstances (Eq. 1) 
calculated for the time of the observation. This is built into our code.
The orbital elements are updated directly by importing the MPC online data,
then ephemeris and the required distances are calculated by Meeus algorithms (Meeus, 1991). 
The sizes of the square windows are defined in km and converted to pixels according to the distance and the pixel scale.
Stars are selected from a list extracted from VIZIER Tycho catalog I/259
(Hog et al., 2000). Hipparcus and Tycho Main catalogue stars are listed by the software. 
$R$ and $I$ magnitudes are extrapolated from V and B magnitudes 
(Caldwell et al., 1993) with different solutions for dwarf and giant stars. 
In the considered colour index range ($0.4 < B-V <0.8$) dwarfs and giants 
extrapolation lead to negligible discrepancies (usually within $0.01$ mag).   
$S$ magnitudes are similarly synthesized by a polynomial solution for M67 
main sequence stars and the
other standard stars of the Vilnius system (Boyle, 1998, Montgomery, 1993).
We note that $R$ or $I$-band excess or departure from $MS$ colors can affect
this step, which error source cannot be totally excluded. The observer should check 
if the determined magnitude of the comparison star is compatible with other filed stars,
with this he can exclude the stars that have deviations of $5$\% or more. 

The final error in $Af\rho$ is calculated as 
$\sigma_{Af\rho}^2 = \sigma_p^2 + \sigma_b^2 + \sigma_n^2 + \sigma_c^2$, where
$\sigma_p$ is the error of coma photometry, $\sigma_b$ is the error introduced by
the sky background subtraction (e.g. Davis, 1990), $\sigma_n^2$ is the readout
noise and $\sigma_c^2$ is the uncertainty of the comparison star brightness. 
The resultant error is typically
between $10$--$20$\%{}. In the worst circumstances (poor $S/N$ of faint comets, which is
on the other hand not the case of Tempel 1) the error may be somewhat higher, indicatively up to 30\%{}.
Data from the same night but from different observers,
who used different comparison stars are very often compatibe.
More specifically, they usually agree within 10\% and rarely disagree more than 20\%.
From the multi-aperture datasets, a normalized $Af\rho$ value is interpolated (or in some cases extrapolated) for a   
5000 km $\rho$ value for each single night and observer. 
Thus the $Af\rho$ data can be compared to the ones 
obtained with photoelectric photometers in previous apparitions with diaphragms of approximately the same size.

\subsection{Observational circumstances}

{\bf Fig 3. comes here}

$Af\rho$ data of comet Tempel 1 is affected by phase effects as the comet was observed in a phase angle range 
$\alpha$ between 11 and 40 degrees. 
The phase corrected $Af\rho$ ($\alpha$=0) can be computed as 
$F_{comet}(\alpha=0)=F_{comet}(\alpha) - C\cdot \alpha$, where $C$ the phase effect coefficient
and is usually between 0.02--0.035 mag/degrees (Meech et al.,1987 -- Lisse et al., 2005) 

Unfortunately, in recent apparitions comet Tempel 1 had very similar close approaches. 
In the case of the recent observations, the geocentric distances around perihelion were similar
as all observed perihelion dates fall within few days in early July. Of course
this means very similar geometric conditions from every points of view.
This makes it impossible to calibrate the phase coefficient 
from these data sets. Unless otherwise indicated, we present uncorrected $Af\rho$ data in the followings.
The other geometric parameter which can influence the activity is the heliocentric
aspect angle of the rotation axis, i.e. the Sun -- comet (center) -- comet north pole angle. 
The bottom panel of Fig 3.
plots its variation, showing that 135 days before perihelion the 
``summer'' of the north pole started. The maximal area which is always illuminated 
refers to the minimal aspect angle 25 days after the perihelion.

\section{Results}

\subsection{General behavior and pre-impact events}

{\bf Fig 4. comes here}

{\bf Fig 5. comes here}

At the beginning of the observing campaign (in late 2004) comet Tempel 1 
was nearly starlike with an image scale around 2000 km/arcsec at 
the comet. $Af\rho$ from this time is also approximative.
In January--February, 2005, an asymmetric coma was observed, 
reaching its largest apparent extension of approximately  $5$' ($75~000$ km) by the approach to our planet 
($0.7$ AU) in early May, 2005. In that period the image scale was $750$ km/arcsec.

Fig 4. compares the $Af\rho$ observed in 2005 and in earlier approaches.
(We must recall that we generally refer to
Afrho values without phase effect corrections, except where explicitally reported.)
The evolution of the comet in 2005 was very similar to those of observed in 1983,
1987, 1994 and 1997--2000 (Starrs et al., 1992, Schleicher et al., 2005, Fink \&{} Hicks, 1996,
Meech et al., 2005, data 
collected by Lisse et al., 2005). The only difference is that the
ascending branch was ``20 days late'', as the same $Af\rho$ values occurred
20 days later with respect to the earlier observations. 

The overall agreement  indicates that Tempel 1 displayed a similar behaviour between 1983 and 2005.
The $Af\rho$ reached its maximum 85 days before the perihelion, later constantly decreased.
A reason of the asymmetric evolution may be a phase effect, but as discussed above, 
an accurate phase coefficient could not be calculated.   
Using a $0.0275$ mag/degree phase coefficient (indicatively an average value among the data reported in the literature) 
the $Af\rho$ behaviour became more symmetric.
After the correction, the $Af\rho$  plateau (around $420$ cm) is more flat and spans between $85$ 
and $4$ days before perihelion.  (Fig. 5 second panel). With this correction the maximum is also consistent with the H$_2$O 
peak production, as reported for previous apparitions (Lisse et al. 2005). This gives a support to the reality 
of the phase effect, too.

The coma profile analysis is presented for the entire apparition in Fig. 6. In the upper panels three profiles (from April 1,
a pre-impact profile from July 3 and a post-impact profile from July 5) are shown. Their log-log fits illustrate
the validity of power-law approximation to this comet and the changes of the coma structure during those 4 months
as compared to the effects of the impact. The evolution of the slopes for the entire apparition is plotted in the bottom 
panel. Between
200--80 days before the perihelion the comet had a moderately compact profile with $d\log Af\rho / d\log \rho
\approx -0.5$, referring to a  $\rho^\gamma$-like surface brightness profile with $\gamma$ typically
around $-1.5$. Later it increased, by about 60 days before perihelion $d\log Af\rho / d\log \rho
\approx -0.5$ was approximatelly $-0.2$. No short-term variations of the logarithmic profile parameter were found.


\subsection{The impact}

{\bf Fig 7. comes here}

{\bf Fig 8. comes here}

The last data points preceding the impact are from images from a remote-controlled telescope
about 20 minutes before the event.
Because of the geographical location of CARA observers (the majority in Europe),
the next point was taken about 15 hours after the impact. 
This enabled us to examine the long-term effects caused by the impact, but it is of course
not enough to deduce very short time-scale effects.

Right before the impact $Af\rho$ decreased very slightly
and d$\log Af\rho$ / d$\log \rho$ was around $-0.1$--$-0.05$. The impact occurred $1.3163$ days before perihelion. 
As $Af\rho$ value would naturally decrease in the comet after the preihelion, the effects of impact
could not be directly characterized from the $Af\rho$ raw curve (Fig 7 upper panel). In order to do this, 
the average trend of the Afrho value was approximated with a simple linear fit to the previous apparitions data, 
from $-25$ to $+25$ days to perihelion (Fig. 7 middle panel). The comparison of the Afrho curves from 
different perihelion passages  allow us to estimate the $Af\rho$ excess resulted by the impact (Fig. 7 bottom panel).
Although this procedure is imprecise in some points (we fitted data of earlier observations then subtracted
from recent observations; all weights were the same in fitting the linear relationship, etc.) this resulted the $\Delta
Af\rho$ of the out-of-impact part having zero average and about 25 cm standard deviation. The total effect of the impact
exceeds the standard deviation with a factor of $6$, supporting the reliability of the conclusions for the
impact effects.

$0.65$ days after the impact the total value of $Af\rho$ had increased up to about $280$ cm,
indicating that much of the dust ejected by the impact had still remained in the coma. Then, the dust excess 
decreased constantly and by $4$--$5$ days after the impact the $Af\rho$ curve is compatible with the regular behaviour
of earlier apparitions, and direct signs of the event were no more detectable.
From the $Af\rho$ excess one can estimate the total dust cross section ejected by the impactor. At
$\rho$=10~000 km, $0.65$, $0.94$ and $1.93$ days after the impact the ``corrected'' $Af\rho$ values were respectively 
$105$ cm, $55$ cm and about $15$ cm above the level of ``normal'' activity referring to the pre-impact values (about 160 cm, 
see the linear regression in Fig. 6). Using Eq. 2., one gets $8.2/A$
km$^2$, $4.3/A$ km$^2$ and $1.2/A$ km$^2$ as the total projected area of dust
grains ($A$ is the albedo).

The impact affected the $Af\rho$--$\rho$ curves, and therefore the 
average profile slope, too. $0.65$ days after the impact the slope $d \log Af\rho/ d \log \rho$ decreased to $-0.32\pm0.01$,
but then relaxed again to $-0.28$, $-0.21$, and $-0.08$ by $1.58$, $1.64$ and $5.61$ days after the impact (Fig 6 bottom panel).
Simultaneously,  on July $4$--$5$ a fan shaped feature within the coma was observed as shown in the panels of Fig. 2. 
The size of both images is 500~000 $\otimes$ 500~000 km, while the inserts are twice magnified
and show the 100~000 $\otimes$ 100~000 km environment around the nucleus. The azimuthal average values are
subtracted from the inserts in order to emphasize the non-radial features (azimutal
renormalization, e.g. Szab\'o et al., 2002). On July 5.85, 1.61 days after the impact the insert
shows this feature on the right side with respect to the solar direction. 4 days later there had remained no
signs of this feature.

This ejecta cloud was also observed in the $Af\rho$ profiles. The last image before impact
showed an almost flat, slightly decreasing profile (Fig 8., top panel). $0.65$ days after the impact
$Af\rho$ has increased significantly everywhere inside the coma, and a cloud (or ``peak'') was observed about
8000 km away from the nucleus. In the following 24 hours this feature moved outward and
broadened slowly (Fig. 8, bottom panel). 
With the assumption that this cloud was produced by the impact, its average projected velocity was calculated to be 0.158 km/s in the first 20 hours
after the impact. Then decelerating to 0.027 km/s about 1.5 days after the event (Fig. 8 top panel). 
The motion of this peak can be interpreted with a "fountain model" of dust cloud propagation (Eddington, 1910, Massonne et al., 1990) 
via a parameter $\beta$, wich is the ratio of repulsive acceleration to solar gravity. The bottom 
panel of Fig 9. shows a fountain-model fit to the measured apparent position of the peak. The initial
projected velocity of the cloud is 0.2 km/s, while $\beta$=0.73$\pm0.04$.

{\bf Fig. 9. comes here}

\subsection{Post-perihelion evolution}

Although the $Af\rho$ is still remained at slightly higher level than before the impact some days after the event, 
it is hard to claim that we still were detecting some effects produced by the impact itself later than $4$--$5$ days after it. 
By this time the impact-specific structures had completely vanished and the coma structure came back to the pre-impact state. 
On the other hand small amplitude Afrho variations cannot be excluded, but if present they are comparable or smaller than the average 
error of our data ($10$--$20$\%). 
$15$ days after the impact the activity did not show any detectable deviation from what happened in previous approaches. 
The only difference being a temporary Afrho increase between $50$--$100$ days after perihelion, which may even be considered as a small 
occasional increment of activity not related to previous events. DI also observed micro-outburtst before the impact, which were not detected by our team, 
as we have no observations 
close enough to the known events. We note that some periods with temporarily increased activity or outbursts
are also present in the post perihelion phase both in 2005 and in previous
approaches, too.

\section{Summary}

The CARA observing campaign allowed us to perform a detailed long term monitoring of comet Tempel 1 during
almost a year of its 2005 apparition. In addition to confirming previously known phenomena associated
with the impact, we derived the conclusions  for a large time span of coverage and also provide
context for quantities during the pre-impact and post-impact periods.
We also discussed the results in the light of published data covering almost 23 years. The major results
can be summarized as follows. The Afrho curve behaved generally just like in previous
oppositions, but the 20-day late of the activity levels in 2005 (as compared to the previous apparitions) has to be mentioned.
Such as previously, in 2005 the $Af\rho$ curve reached the maximal value $83$ days before perihelion 
($Af\rho = 310$ cm). In 2005, the ascending branch of $Af\rho$ curve was about $20$ days late compared to earlier apparitions. 
This can at least partially be a phase effect. The log-slope of $Af\rho$ was around $-0.5$ between about 180--100 days
before the impact but then
evolved near the steady-state like $0$ value by the impact time.
The effects of the impact were clearly detected in the first $3$ days.
Otherwise, just before and ``after'' the impact, the coma profile was almost flat, and  the impact  did not produce
detectable permanent effects in the slope. 
The impact resulted at least $60$\% increase of $Af\rho$, and in 4--5 July an extending dust cloud was observed.
Its projected motion was consistent with the fountain model with an inner speed of $0.2$ km/sec  
and $\beta = 0.027 \pm 0.04.$ After the impact the dust cross section temporary  increased to 8.3/$A$ km$^2$ 
for $\rho$ = $10~000$ km, respecting to the pre-impact state, the excess then relaxed
in about 2 days.

\acknowledgements{\bf Acknowledgements.}
{
The support of the Hungarian OTKA Grant T 042509 is acknowledged. GyMSz was supported by the
Magyary Zoltán Higher Educational Public Foundation. The contribution of M. Fulle, G.P. Tozzi, L. 
Jorda and members of the Union of Italian 
Amateur Astronomers (U.A.I. http//cara.uai.it) helped develop the project. 
We thank all the contributing observers listed in Table 1. the indispensable assistance in measurements.
We also thank 
M. Barbieri
(CISAS - Dept. Physics of Padova University) for useful suggestions and
assistance.
}

\begin{center}
\begin{table*}
\begin{tabular}{llllll}
\hline
Name& Observatory& Country& Res.& Filter\\
     &            &        & \footnotesize arcs/pix& &\\
\hline
Aletti A.,Buzzi L. & Schiaparelli (Varese) & Italy & 1.5 & BVRI\\
Bryssinck E. & Brixis (Kuibeke) & Belgium & 2.05& V\\
Buso V.,Mazalan V.& Colegio Cristo Rei (Rosario)& Argentina& 1.9& RI\\
Carosati D.& Armenzano (Assisi)& Italy & 1.63& VRI\\
Focardi L. & private site (Florence) & Italy & 2.57& unf.\\
Fratev F. & private site (Milan) & Italy & 1.47& unf.\\
Guido E. & \footnotesize New Mexico Skies (Sacramento hills) &NM, USA & 
2.15 & VRI\\
Higgins D.&  Hunters Hills (Canberra) & Australia&  1.32&  RI\\
Ligustri R. & \vline \\
\small Romanello F., Da Rio D.& \vline \ Talmassons  (Udine) &Italy 
&   2.39 & BVR,Rc\\
Mikuz H.& Crni Vrh (Crni Vrh)& Slovenia& 2.5&  VR\\
Milani G.& private site (Padova) & Italy & 2.3& RI \\
Nicolini M. & Cavezzo (Modena) & Italy & 2.24&  R,Rc\\
Scarmato T. & S. Costantino di Briatico (VV) & Italy & 2.6&  R\\
\small Sostero G., Gonano M., & \vline\\
\small Gonano V., Lepardo A., & \vline \ Remanzacco (Udine) & Italy & 
several$^{*}$ & BVRI,Rc \\
Santini V. & \vline\\
Tirelli D & Tirelli (Sossano)  & Italy & 1.2 & R\\
Trabatti R. & Descartes (Chignolo Po) & Italy &  2.3 & RI,Rc\\
Zattera F.  & private site (Malo) & Italy &  2.8 &  BV\\
\hline
\end{tabular}

$^{*}$ 2.15, 1.9 - 3.48 - 3.8, 2.0\\
SBIG ST8 XE, SXV-M7, Hi-Sis 23 ME
\caption{Contributing observers and observing sites.
filter codes are: B, V, R, I Johnson-Cousins filters, Vilnius S and Rc:
647/10 nm red continuum.}
\end{table*}
\end{center}

\newpage

Figure captions:

Fig 1.: Spectrum of a comet with superimposed bandpasses of the filters

Fig 2.: {Comet 9P/Tempel as observed on July 5.85 and 9.84 UT at Talmassons
Observatory. 
The inserts show azimuthally renormalized images (e.g. Szab\'o et al., 2002). 
The inserts are twice magnified and their size is 100~000$\times$100~000km.
Note the twisted features in the coma after the impact.}

Fig 3.: {The geometric circumstances. Top panel:
solar phase angle in past oppositions; dotted line -- 1983; dashed line --
1994; solid line -- 2005. Bottom panel: The change of the solar aspect angle
(Sun -- comet -- north pole) in 2005. Pole coordinates are from DI
imaging (solid line) and photometry (dotted line).
}

Fig 4.: {CARA $Af\rho$ data (black circles) superimposed to
previous oppositions. Squares: 1983, 1994, 1997--2000, (collected by Lisse et al., 2005).}

Fig 5.: {CARA data for 9P/Tempel extrapolated to $\rho=5000$ km. Top
panel: without solar phase correction, middle panel: with 0.0275 mag/degree correction applied.
Bottom panel: the evolution of the logarithmic slope during the 2004--2005 apparition.
Filter codes are: dots -- R; triangles -- I; open circles -- S; $Af\rho$ plots are log-scaled.
}

Fig 6.: {Log-log scaled profiles of 9P/tempel 1 on selected nights: 1 April,
3 July and 5 July, from top to third panel, respectively.
Bottom panel: the evolution of the logarithmic slope during the 2004--2005 apparition.}

Fig 7.: {Top panel: $Af\rho$ around the impact. Second panel: evolution of
$Af\rho$ at the same time in previous oppositions (Lisse et al., 2005). 
Third panel: the difference 
regarding to the linear fit is considered to show the effects of the impact.
Bottom panel: Evolution of the logarithmic slope of the coma.}

Fig 8.: {Top panel:
Evolution of $Af\rho$ vs. rho. The top panel represents the unaffected activity $\tau=0.01$ days 
before (triangles) and $5.62$ days after the impact (dots). Bottom panel: the
evolution of the coma $0.65$, $0.94$ and $1.93$ days after the impact.}

Fig 9.: {Propagation of the peak in Fig. 7. Top: the average velocities in
different sections. Bottom: a fountain model solution is fitted with $v=0.2$ km/s, $\beta=0.73$.}

\begin{figure}[h]
\centering\includegraphics[bb=58 85 384 260, width=12cm]{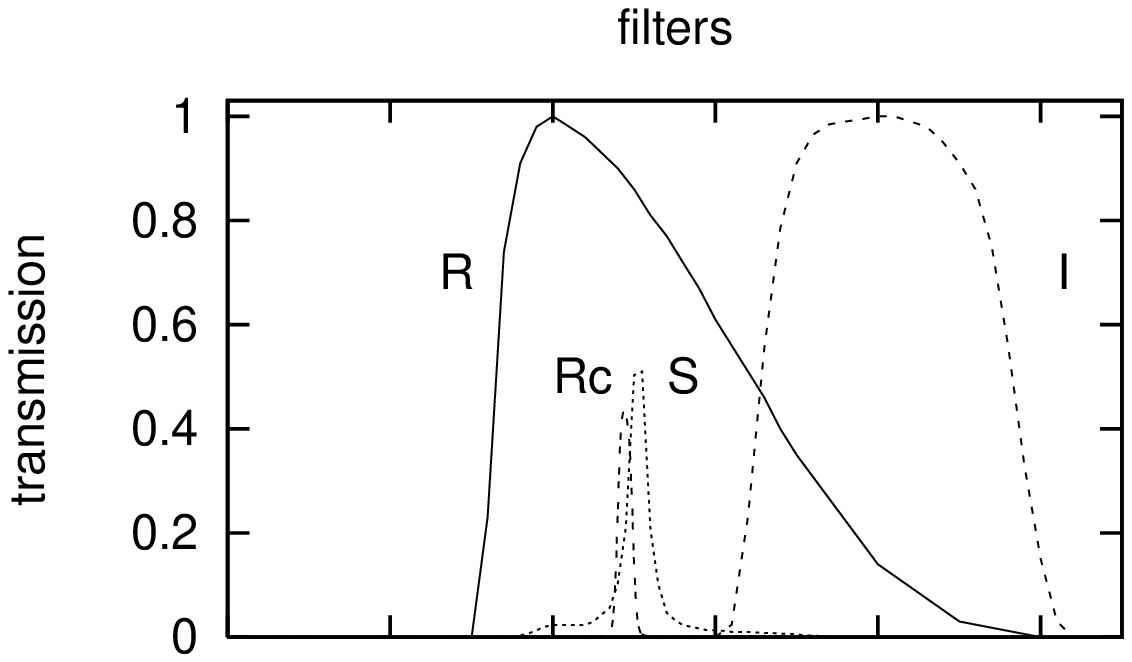}
\centering\includegraphics[bb=58 52 384 178, width=12cm]{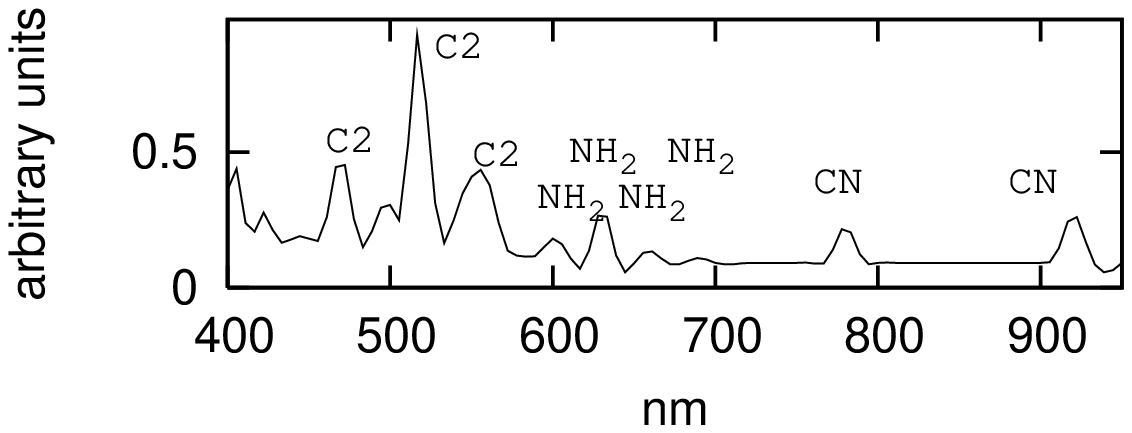}
\caption{Spectrum of a comet with superimposed bandpasses of the filters}
\end{figure}

\newpage

\begin{figure}[h]
\centering\includegraphics[bb=20 20 575 288, width=14cm]{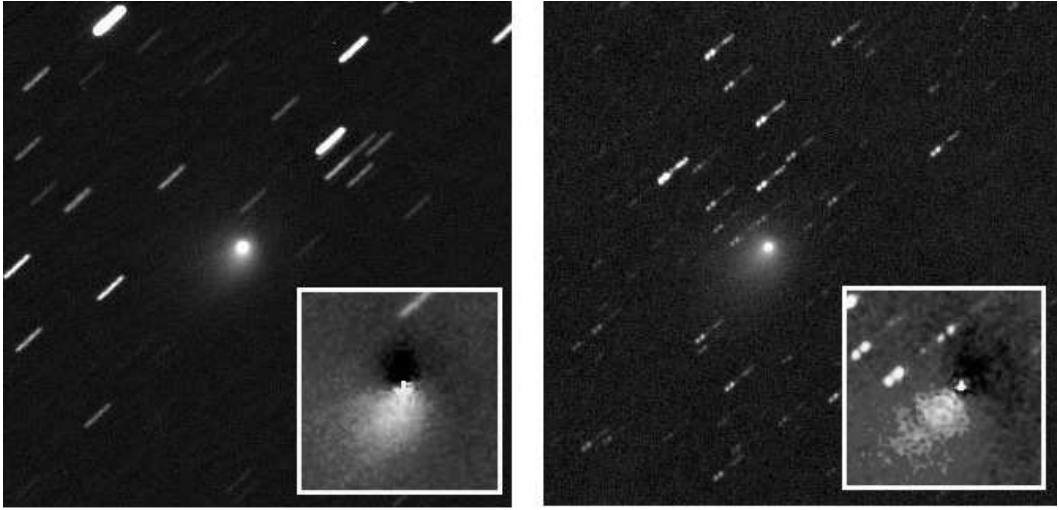}
\caption{Comet 9P/Tempel as observed on July 5.85 and 9.84 UT at Talmassons
Observatory. 
The inserts show azimuthally renormalized images (e.g. Szab\'o et al., 2002). 
The inserts are twice magnified and their size is 100~000$\times$100~000km.
Note the twisted features in the coma after the impact.}
\end{figure}

\begin{figure}[h]
\centering\includegraphics[bb=58 80 384 226, width=12cm]{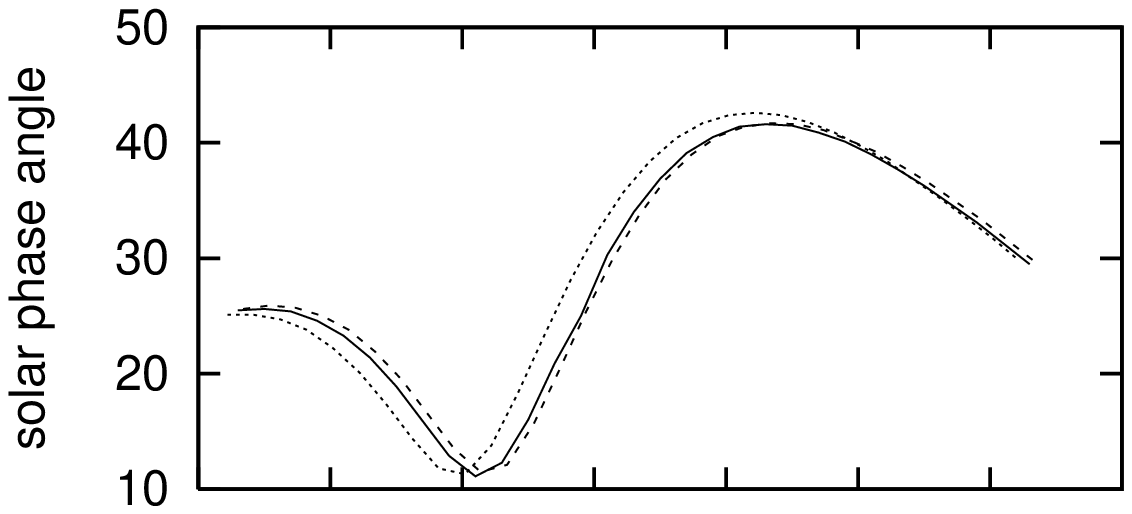}
\centering\includegraphics[bb=58 52 384 235, width=12cm]{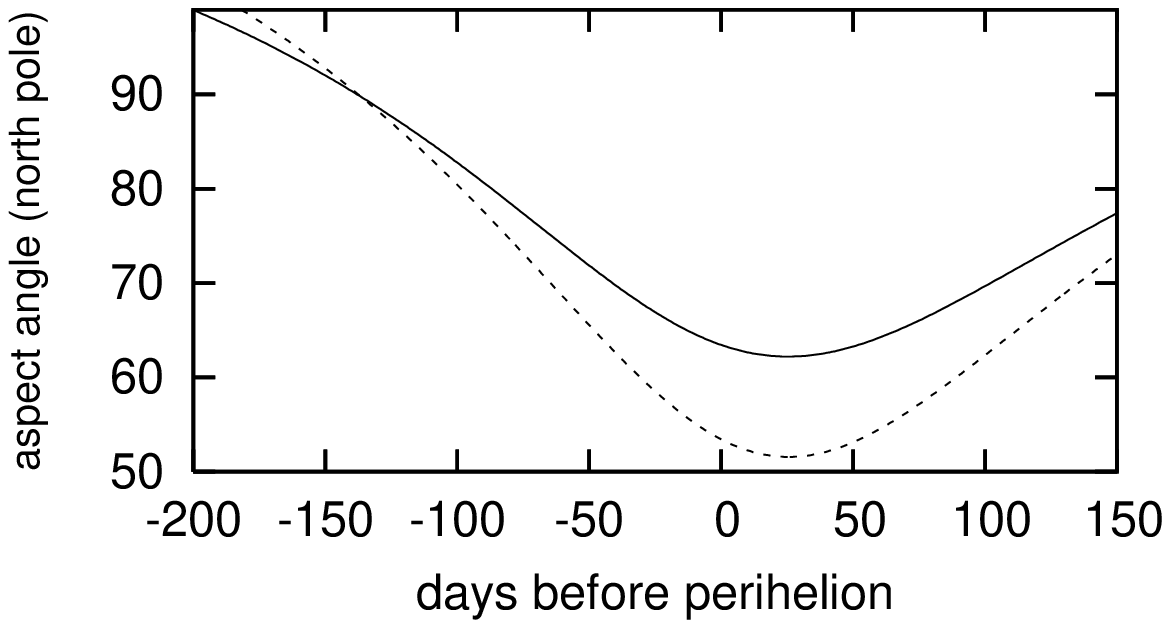}
\caption{The geometric circumstances. Top panel:
solar phase angle in past oppositions; dotted line -- 1983; dashed line --
1994; solid line -- 2005. Bottom panel: The change of the solar aspect angle
(Sun -- comet -- north pole) in 2005. Pole coordinates are from DI
imaging (solid line) and photometry (dotted line).
}
\end{figure}

\begin{figure}[h]
\centering\includegraphics[bb=58 52 384 350, width=12cm]{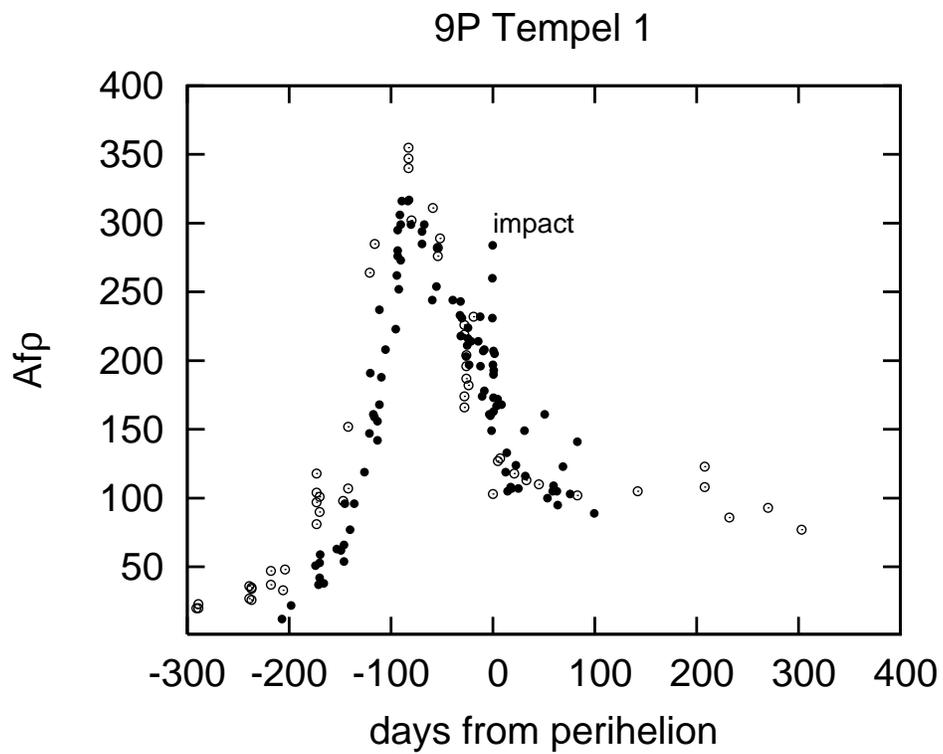}
\caption{CARA $Af\rho$ data (black circles) superimposed to
previous oppositions. Squares: 1983, 1994, 1997--2000, (collected by Lisse et al., 2005).}
\end{figure}

\begin{figure}[h]
\centering\includegraphics[bb=58 86 384 230, width=12cm]{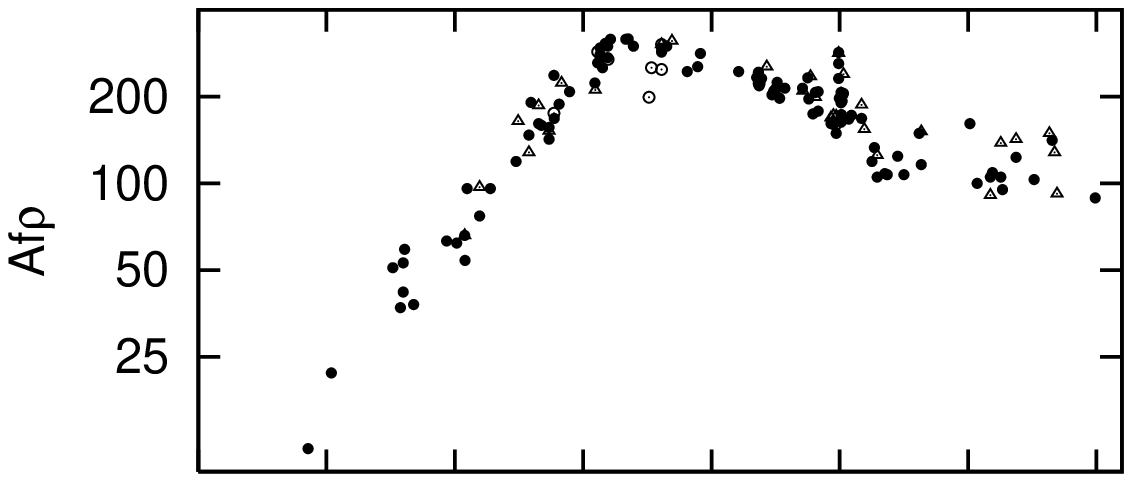}
\centering\includegraphics[bb=58 50 384 230, width=12cm]{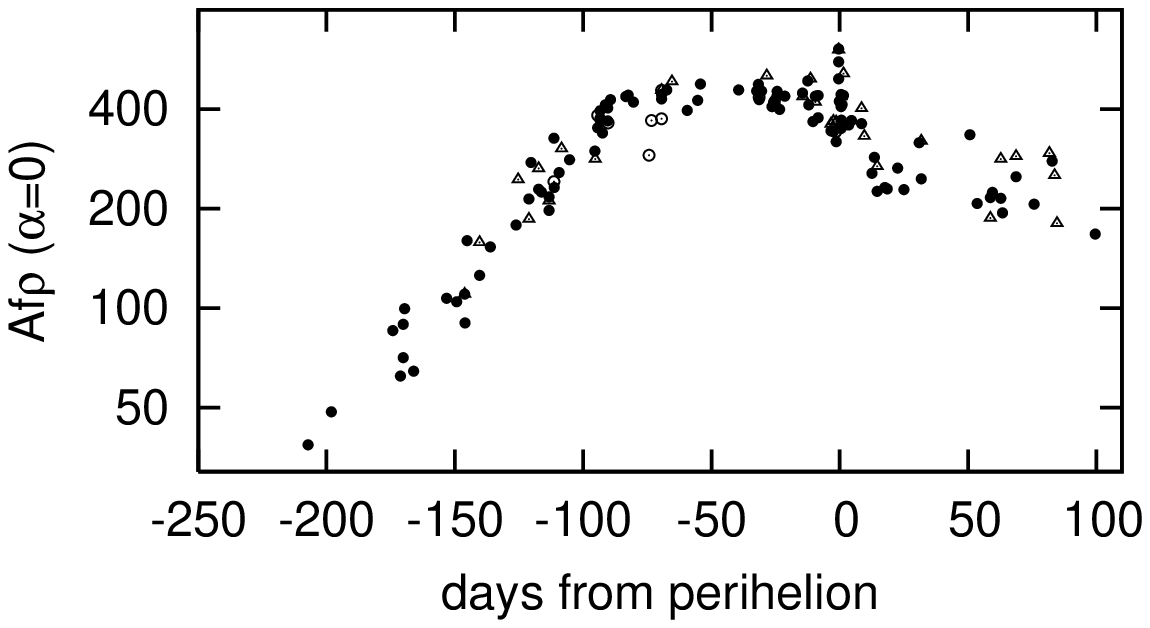}
\caption{{CARA data for 9P/Tempel 1 extrapolated to $\rho=5000$ km. Top
panel: without solar phase correction, middle panel: with 0.0275 mag/degree correction applied.
Filter codes are: dots -- R; triangles -- I; open circles -- S}}
\end{figure}

\begin{figure}[h]
\centering\includegraphics[bb=58 80 384 165, width=12cm]{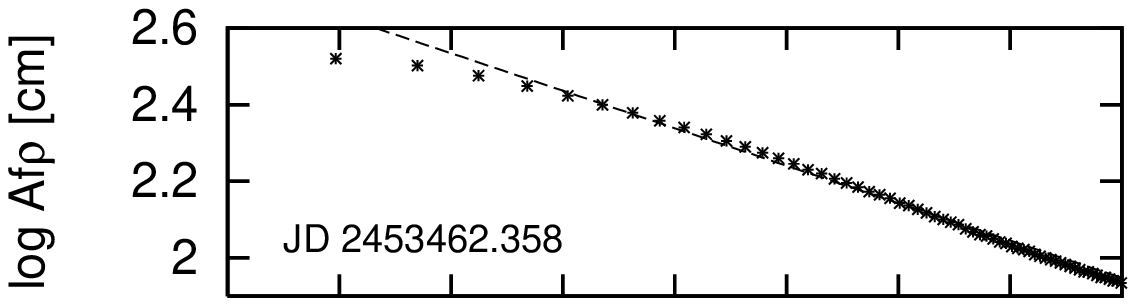}
\centering\includegraphics[bb=58 75 384 165, width=12cm]{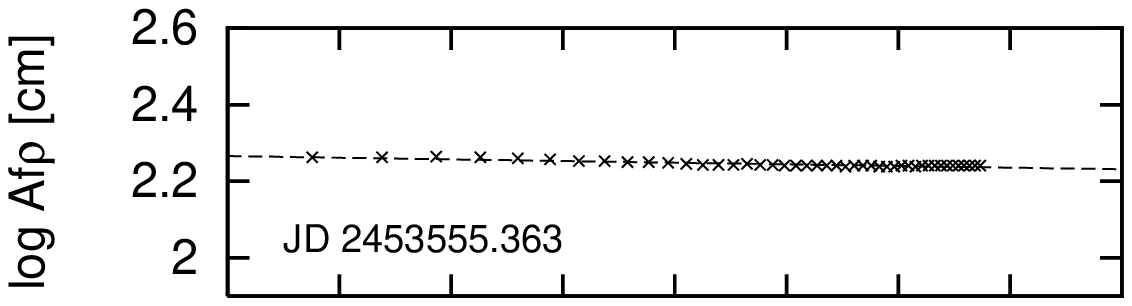}
\centering\includegraphics[bb=58 52 384 175, width=12cm]{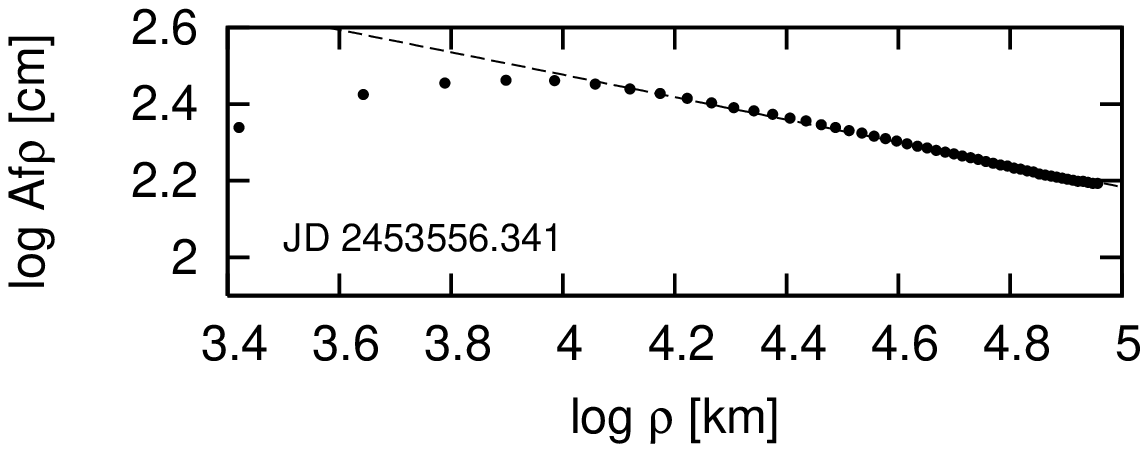}
\centering\includegraphics[bb=60 52 384 178, width=12cm]{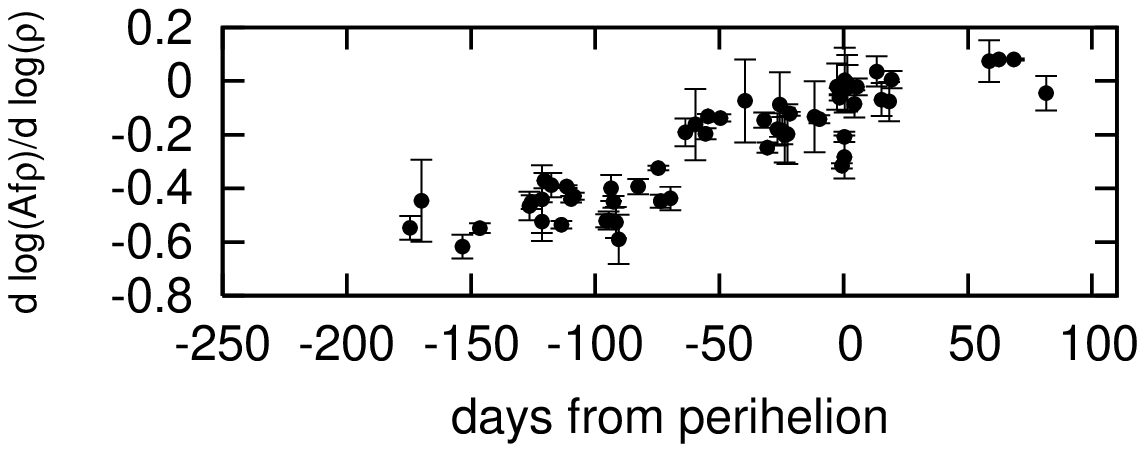}
\caption{Log-log scaled profiles of 9P/tempel 1 on selected nights: 1 April,
3 July and 5 July, from top to third panel, respectively.
Bottom panel: the evolution of the logarithmic slope during the 2004--2005 apparition.}
\end{figure}

\begin{figure}[h]
\centering\includegraphics[bb=50 87 400 178, width=12cm]{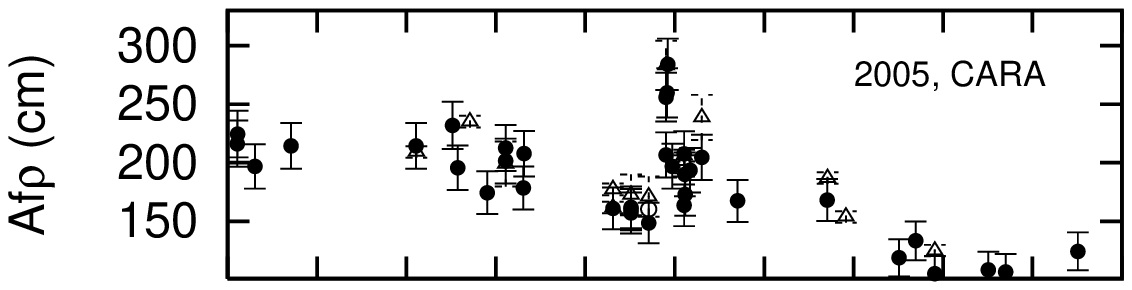}
\centering\includegraphics[bb=52 87 389 178, width=12cm]{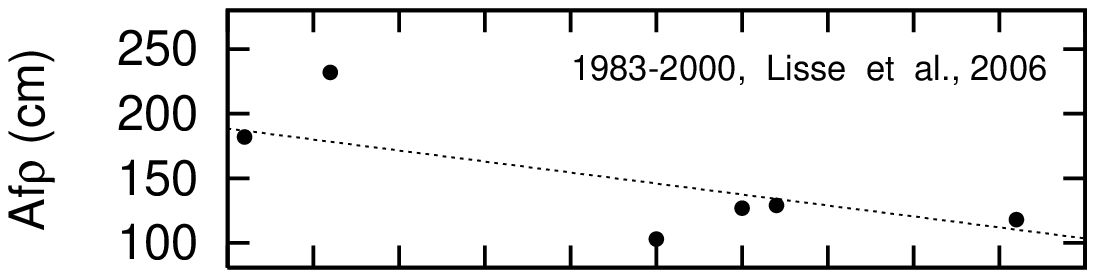}
\centering\includegraphics[bb=50 87 400 178, width=12cm]{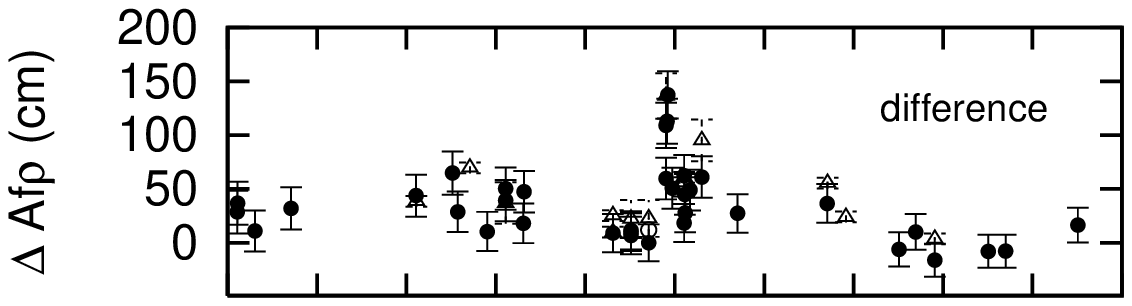}
\centering\includegraphics[bb=50 52 417 180, width=12cm]{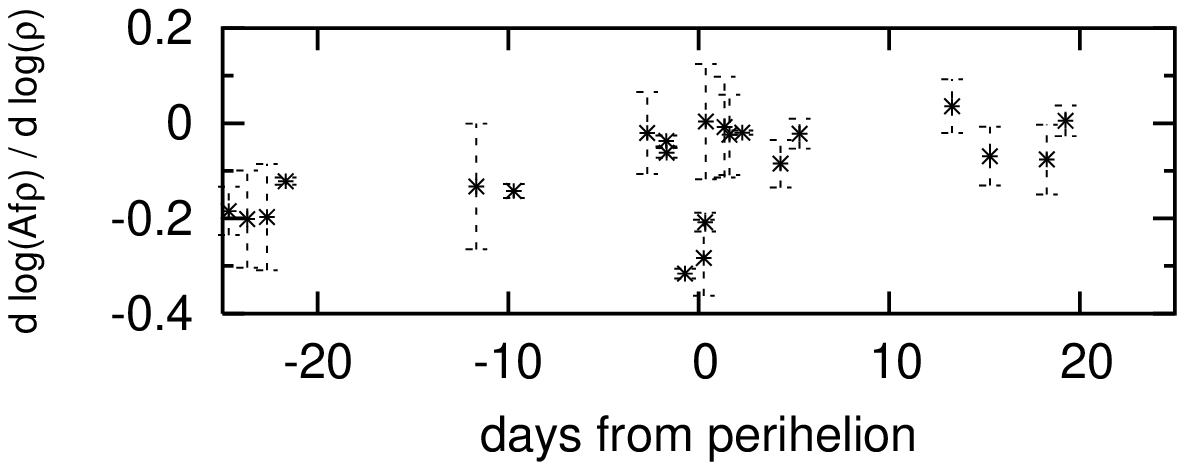}
\caption{
{Top panel: $Af\rho$ around the impact. Second panel: evolution of
$Af\rho$ at the same time in previous oppositions (Lisse et al., 2005). 
Third panel: the difference 
regarding to the linear fit is considered to show the effects of the impact.
Bottom panel: Evolution of the logarithmic slope of the coma.}
}
\end{figure}

\begin{figure}[h]
\centering\includegraphics[bb=52 89 385 150, width=12cm]{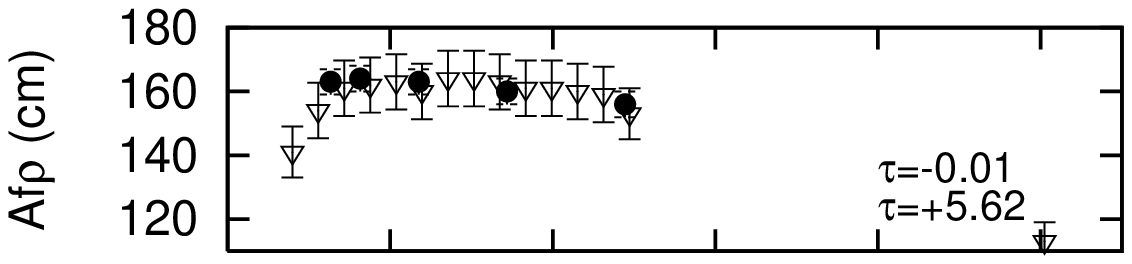}
\centering\includegraphics[bb=52 52 385 290, width=12cm]{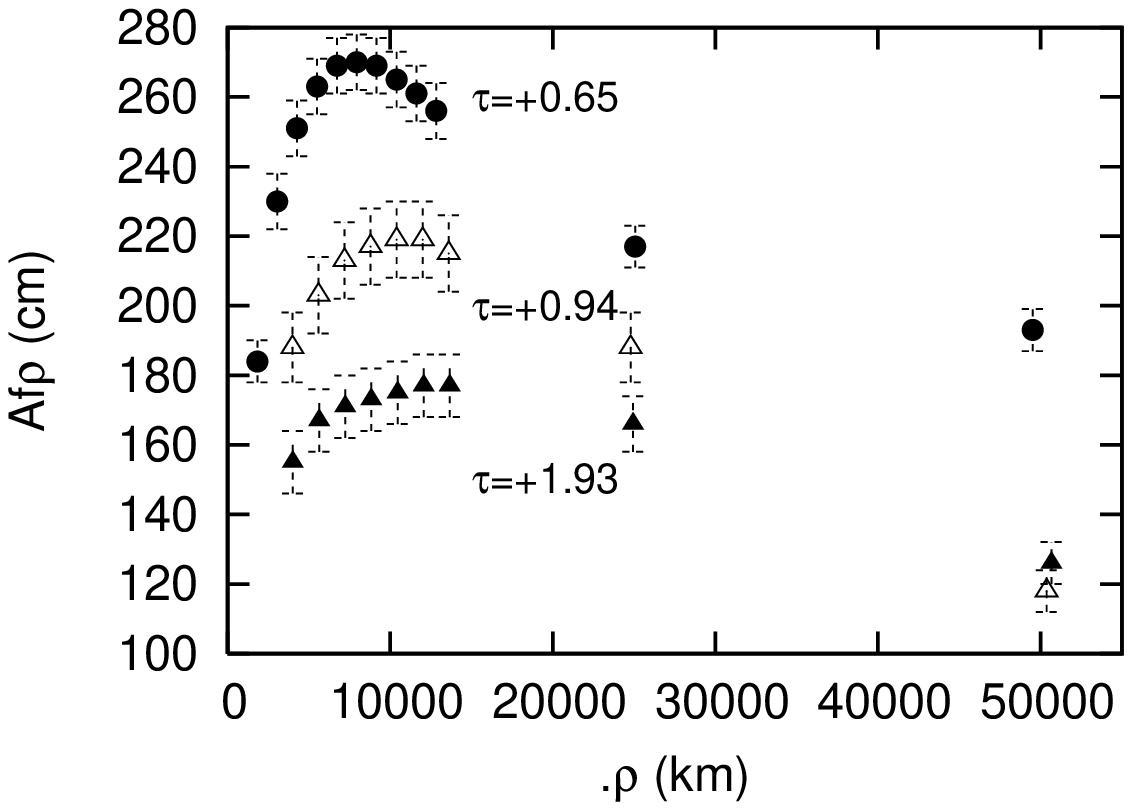}
\caption{
{Top panel:
Evolution of $Af\rho$ vs. rho. The top panel represents the unaffected activity $\tau=0.01$ days 
before (triangles) and $5.62$ days after the impact (dots). Bottom panel: the
evolution of the coma $0.65$, $0.94$ and $1.93$ days after the impact.}
}
\end{figure}

\begin{figure}[h]
\centering\includegraphics[bb=50 80 384 230, width=12cm]{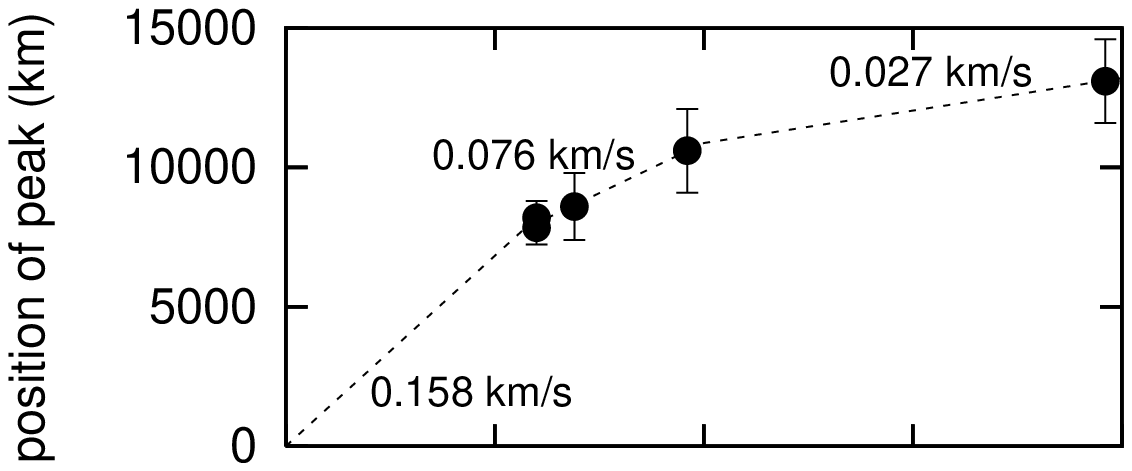}
\centering\includegraphics[bb=50 50 384 230, width=12cm]{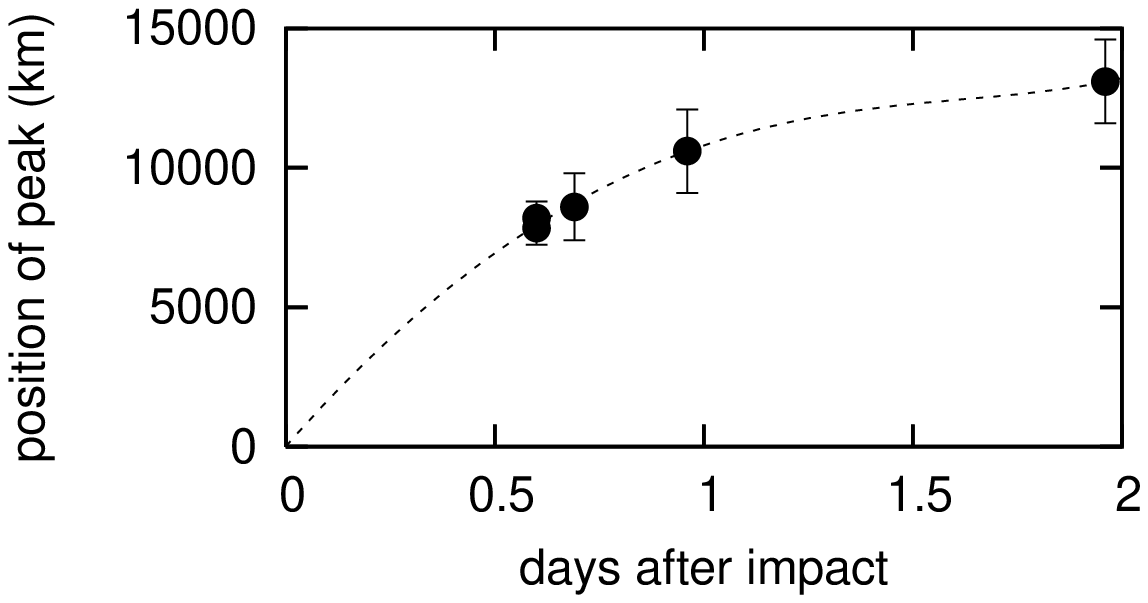}
\caption{
{Propagation of the peak in Fig. 7. Top: the average velocities in
different sections. Bottom: a fountain model solution is fitted with $v=0.2$ km/s, $\beta=0.73$.}
}
\end{figure}

\end{document}